\documentclass[USenglish]{lipics-v2016}
\usepackage{enumitem}
\usepackage{siunitx}
\usepackage{subcaption}
\usepackage{amsmath}
\newtheorem{proposition}{Proposition}

\usepackage{algorithm}
\usepackage{algorithmic}

\usepackage{tikz}
\usetikzlibrary{positioning,calc}
\tikzstyle{vertex}=[auto=left,circle,draw=black,fill=black!25,minimum size=20pt,inner sep=0pt]
\tikzstyle{svertex}=[auto=left,circle,draw=black,fill=black,minimum size=7pt,inner sep=0pt]

\usepackage{xcolor}
\definecolor{goodgreen}{HTML}{7FC97F}
\definecolor{goodpurple}{HTML}{BEAED4}
\definecolor{goodorange}{HTML}{FDC086}
\definecolor{goodyellow}{HTML}{FFFF99}
\definecolor{goodblue}{HTML}{386CB0}

\usepackage{float}

\usepackage{xspace}
\newcommand{\concuss}{\textsf{CONCUSS}\xspace}
\newcommand{\nxvf}{\textsf{NXVF2}\xspace}

\newcommand{\Dvorak}{Dvo{\v{r}}{\'a}k }

\newcommand{\naive}{na\"{i}ve }

\author{Michael P.~O'Brien}
\author{Blair D.~Sullivan}
\affil{North Carolina State University, \texttt{(mpobrie3|blair\_sullivan)@ncsu.edu}}
\title{An Experimental Evaluation of a Bounded Expansion Algorithmic Pipeline}

\begin{document}

    \maketitle
    \begin{abstract}
    	Previous work has suggested that the structural restrictions of graphs from classes of \emph{bounded expansion}\textemdash locally dense pockets in a globally sparse graph\textemdash naturally coincide with common properties of real-world networks such as clustering and heavy-tailed degree distributions.
As such, fixed-parameter tractable algorithms for bounded expansion classes may offer a promising framework for network analysis where other approaches have struggled to scale.
However, there has been little work done in implementing and evaluating the performance of these structure-based algorithms.
To this end we introduce \concuss, a proof-of-concept implementation of a generic algorithmic pipeline for classes of bounded expansion.
In particular, we focus on using \concuss for \emph{subgraph isomorphism counting} (also called motif or graphlet counting), which has been used extensively as a tool for analyzing biological and social networks.
Through a broad set of experiments we first evaluate the interactions between implementation/engineering choices at multiple stages of the pipeline and their effects on overall run time.
From there, we establish viability of the bounded expansion framework by demonstrating that in some scenarios \concuss achieves run times competitive with a popular algorithm for subgraph isomorphism counting that does not exploit graph structure.
Finally, we empirically identify two particular ways in which future theoretical advances could alleviate bottlenecks in the algorithmic pipeline.

    \end{abstract}
    \section{Introduction}
The analysis of \emph{complex networks} has proven useful for understanding relationships between entities in large data sets.
For example, networks modeling social interactions, protein functionality, and accessible transportation routes have been studied extensively in the network science community.
However, since many network properties that might yield interesting insights are NP-hard to compute, the prevailing strategies for detecting such properties generally depend on sampling, estimation, and/or heuristics whose reasons for success or failure are not theoretically well understood.

At the same time, the structural graph theory and parameterized algorithms communities have a large body of literature on exactly solving NP-hard problems in polynomial time on sparse graphs by exploiting the underlying graph structure.
Unfortunately, many of these results either rely on structure like low treewidth, which is unlikely to occur in many complex networks, or hide enormous constants in big-O notation that in practice negate the theoretical gains from the polynomial run times.
Recent work has identified a structural class of graphs known as \emph{bounded expansion} that gives significant algorithmic advantages~\cite{boundedexpansion1,boundedexpansion2,dvorak2009algorithms} while accommodating the structural characteristics of complex networks~\cite{demaine2014structural}.
Though there are some algorithms for bounded expansion that appear to be suitable for practical use, they have yet to be implemented and tested in order to understand their behavior.

In this work, we seek to establish the use of bounded expansion structure as a viable pathway to creating scalable software tools for network analysis.
As a starting point, we focus specifically on the problem of \emph{subgraph isomorphism counting} in which the number of occurrences of a small subgraph $H$ (sometimes called a \emph{motif} or \emph{graphlet}) in a larger host graph $G$ is counted.
Because previous research suggests that subgraph isomorphism counting can be used as a building block for understanding network dynamics~\cite{shen2002network} and aligning similar graphs~\cite{milenkovic2008uncovering}, an improved counting algorithm would have many practical applications.
Our main contribution is a set of computational experiments using \concuss~\cite{concuss}, an open source software tool written in pure Python that implements the best known algorithm for subgraph isomorphism counting in bounded expansion graphs~\cite{demaine2014structural}.
On a high level, \concuss finds a coloring of the vertices to identify the relevant structure, uses the coloring to decompose the graph into small pieces, counts the isomorphisms on each small piece, and combines the subsolutions to get a final count.
Notably, the theoretical description of this pipeline leaves room for multiple choices for implementing certain subroutines; we implement several variants of these routines and discuss the associated algorithm engineering decisions.

Our experimental design is oriented towards understanding not only the strengths of the algorithmic pipeline of \concuss but also the ways in which further theoretical research could alleviate bottlenecks.
Theoretical advances to this pipeline would not only enable faster subgraph isomorphism counting, but also provide an improved framework for solving other combinatorial problems in graphs of bounded expansion (see Section~\ref{sec:concuss}).
As such, we answer the following four questions:
\begin{enumerate}[label=Q\arabic*.]
	\item What choices of subroutines give the best performance in \concuss?
	\item How does \concuss compare to other algorithms for subgraph isomorphism counting?
	\item How does the distribution of color class sizes influence downstream performance?
	\item Would alternative colorings with weaker structural guarantees incur significant penalties?
\end{enumerate}

After giving the necessary background and notation in Section 2, we briefly describe the algorithmic pipeline \concuss implements in Section 3.
Section~\ref{sec:design} details the general experimental setup.
In Section~\ref{sec:q12} we show the performance of various configurations of \concuss compared to each other (Q1) and to the NetworkX~\cite{networkx} implementation of the popular VF2 algorithm~\cite{cordella2004sub} (Q2).
We then investigate in Section~\ref{sec:q34} ways in which theoretical improvements to the coloring algorithms might improve \concuss (Q3 and Q4).

    \section{Background}
All graphs in this paper are assumed to be simple and undirected unless otherwise specified.
For a graph $G$, let $V(G)$ and $E(G)$ denote the vertices and edges of $G$, respectively, and let $uv$ denote an edge with endpoints $u$ and $v$.
For convenience, we define $|G|=|V(G)|$.
We will write $P_n$ and $S_n$ for the path and star on $n$ vertices, respectively.
A \emph{(vertex) coloring} $\phi$ is a mapping of $V(G)$ to a set of $k$ colors; we say that $k$ is the \emph{size} of $\phi$.
A \emph{color class} $C$ of $\phi$ is a maximal set of vertices that for all $u,v\in C$, $\phi(u) = \phi(v)$.

The remainder of this section gives background on subgraph isomorphism counting, bounded expansion, and a related structural class called \emph{bounded treedepth}.

\subsection{Subgraph Isomorphism Counting}
\begin{definition}
	Given graphs $G,H$, a \emph{subgraph isomorphism} from $H$ to $G$ is a mapping $\psi: V(H) \to V(G)$ such that $uv\in E(H) \iff \psi(u)\psi(v)\in E(G)$.
\end{definition}
The subgraph isomorphism counting problem is to count the number of distinct isomorphisms from $V(H)$ to $V(G)$; solving this problem is \#W[1]-hard\footnote{{There is likely no $O(f(|H|)\cdot n^c)$ time algorithm for any computable function $f$ and constant $c$.}}~\cite{jerrum2013parameterised}.
The most successful algorithms to date used for subgraph isomorphism counting have relied on backtracking~\cite{ullmann1976algorithm,bonnici2013subgraph} or expanding partial mappings~\cite{cordella1999performance,cordella2004sub,carletti2017introducing}.
Other subclasses of algorithms have focused on indexing graph databases~\cite{shang2008taming,he2008graphs} or creating data structures for parallel computation~\cite{sun2012efficient}.
\looseness-1

Subgraph isomorphism counting is important as a building block for two other network analysis techniques, described below.

\textit{Motif Counting}:
Originally used in the study of cellular biology~\cite{shen2002network}, \emph{motif counting} computes the number subgraph isomorphisms expected to occur ``by chance'' and compares it to the observed subgraph isomorphism count.
If $H$ represents some functional unit in the domain of the data\textemdash i.e., cliques in social networks corresponding to people with common interests\textemdash an unexpected abundance of $H$ in $G$ gives insight into the domain dynamics that caused these motifs to occur.

\textit{Graphlet Degree Distribution}:
Subgraph isomorphism counting is also a building block in the \emph{graphlet degree distribution}~\cite{prvzulj2007biological}, which measures the number of times a vertex occurs in distinct ``positions'' of multiple small subgraphs.
Graphlet degree distributions give a more robust ``fingerprint'' of a graph's structure which has in turn been used to uncover underlying domain knowledge~\cite{milenkovic2008uncovering}.
They also have been used in network alignment~\cite{milenkovic2010optimal}, using the knowledge that two vertices are more likely to be mapped to one another if they occupy the same positions at similar frequencies.

\subsection{Bounded Expansion}
The algorithms implemented in \concuss are designed to exploit \emph{bounded expansion} network structure.
While classes of bounded expansion can be characterized by their $r$-shallow minors~\cite{boundedexpansion1} or their neighborhood complexities~\cite{reidl2016characterising}, we focus on the algorithmically relevant characterization using \emph{$p$-centered colorings}~\cite{boundedexpansion1}.

\begin{definition}\label{def:p_centered_coloring}
	A \emph{$p$-centered coloring} of graph $G$ is a vertex coloring such that every connected subgraph $H$ has a unique color or uses at least $p$ colors.
	The minimum size of a $p$-centered coloring of $G$ is denoted $\chi_p(G)$.\looseness-1
\end{definition}
\begin{proposition}\cite{boundedexpansion1}\label{prop:pcc_be}
	 A class of graphs $\mathcal{C}$ has bounded expansion if and only if for some function $f$, every graph $G\in \mathcal{C}$ satisfies $\chi_p(G) \leq f(p)$ for all $p\geq 1$.
\end{proposition}
Intuitively, $p$-centered colorings identify overlapping, locally well-structured regions of the graph; being in a class of bounded expansion implies the graph can be covered by a small number of such regions.
More specifically we note that by Definition~\ref{def:p_centered_coloring}, for any $p$-centered coloring $\phi$ of graph $G$, if there is a subgraph $H$ for which $\phi|_H$ uses fewer than $p$ colors, then every subgraph of $H$ has a vertex of unique color.
This relates classes of bounded expansion to a much more restricted class of graphs described below.

\subsection{Centered Colorings and Treedepth}
\begin{definition}\label{def:centered_coloring}
	A \emph{centered coloring} of graph $G$ is a vertex coloring such that every connected subgraph $H$ has a vertex of unique color, called a \emph{center}.
	The minimum number of colors needed for a centered coloring of $G$ is its \emph{centered coloring number}, denoted $\chi_{\mathrm{cen}}(G)$.
\end{definition}
Note that a centered coloring is also proper, or else there would be a connected subgraph of size two with no center.
Centered colorings are closely related to \emph{treedepth decompositions}.

\begin{definition}\label{def:treedepth}
	A \emph{treedepth decomposition} of a graph $G$ is an injective mapping $\psi: V(G) \to V(F)$, where $F$ is a rooted forest and $uv \in E(G)\implies \psi(u)$ is an ancestor or descendant of $\psi(v)$.
	The \emph{depth} of a treedepth decomposition is the height of $F$.
	The \emph{treedepth} of $G$, denoted $\operatorname{td}(G)$ is the minimum depth of a treedepth decomposition of $G$.
\end{definition}
In other words, a treedepth decomposition arranges the vertices of $G$ in such a way that no edge joins vertices from different branches of the tree.
Given a centered coloring with $k$ colors, we can generate a treedepth decomposition of depth at most $k$ by choosing center $v$ to be the root and recursing on the components of $G\backslash \{ v\}$, as detailed in Algorithm~\ref{alg:cc_to_tdd} (Appendix~\ref{app:tdd}).
In this way, Definition~\ref{def:p_centered_coloring} is equivalent to the fact that every small set of colors in a $p$-centered coloring induces components with small treedepth.

    \section{CONCUSS}\label{sec:concuss}
Broadly speaking, the algorithmic pipeline in \concuss executes four modules
 to count the number of isomorphisms of graph $H$ in $G$.
\begin{enumerate}
	\item \textsc{Color}: \\
   Find a $(|H|+1)$-centered coloring of $G$.
	\item \textsc{Decompose}: \\
  Compute a treedepth decomposition of the subgraph induced by each set of $|H|$ colors.
	\item \textsc{Compute}: \\
   Using dynamic programming, count the isomorphisms in the treedepth decompositions.
	\item \textsc{Combine}: \\
   Combine the counts from previous step to get the total number of isomorphisms in $G$.
\end{enumerate}

In Appendix~\ref{app:concuss}, we briefly describe these modules and the algorithm engineering aspects considered as part of the implementation.
Many of the subroutines in the modules can achieve the same high-level functionality with different implementation details, e.g., heuristics, data structures, etc.
We draw particular attention to the following areas of algorithm engineering:
(1) post-processing the $p$-centered coloring to merge color classes and reduce the coloring size (\textsc{Color}), (2) saving partial treedepth decompositions when enumerating sets of colors (\textsc{Decompose}), and (3) representing partial isomorphisms as bitvectors to enable fast checking of consistency (\textsc{Compute}).
\concuss allows users to specify their choices for subroutines in a configuration file; for more details, see the \concuss source and documentation~\cite{concuss}.
These alternatives are designated with \texttt{teletype}.\looseness-1

We note that the four-stage algorithmic workflow (\textsc{Color}, \textsc{Decompose}, \textsc{Compute}, \textsc{Combine}) can be useful for solving problems other than subgraph isomorphism counting.
In particular, this approach is amenable to problems in which local subsolutions can be efficiently computed on graphs of low treedepth and then combined to create a global solution.
For such a problem, it is possible to reuse the \textsc{Color} and \textsc{Decompose} modules and simply replace the \textsc{Compute} and \textsc{Combine} modules with the appropriate alternatives.
\Dvorak et al.~\cite{dvorak2013testing} proved that any problem expressible in first-order logic can be solved in this way, such as finding small dominating sets or small independent sets.

    \section{Experimental Design}\label{sec:design}
We conducted four experiments, one to answer each question posed in Section 1.
The first two (Section~\ref{sec:q12}) identify an ``optimal'' configuration for \concuss and test its scalability against an existing subgraph isomorphism counting algorithm.
The other two experiments (Section~\ref{sec:q34}) investigate ways in which properties of the coloring affect downstream computation to guide future theoretical research.

To compare the effects of varying multiple options, we normalize two measures $a,b$ using the difference to sum ratio, i.e., $\frac{a-b}{a+b}$.
This ratio will be close to $1$ when $a\gg b$ and close to $-1$ when $b\gg a$.
When comparing times, a positive ratio indicates $b$ is faster than $a$.

\subsection{Data}\label{sec:rgm}
Since the theoretical advantages of the algorithms implemented in \concuss rely on exploiting structure, it was necessary to ensure our input graphs belonged to classes of bounded expansion.
Consequently, we chose three random graph models which asymptotically almost surely produce classes of bounded expansion~\cite{demaine2014structural}: the stochastic block model (SB)~\cite{holland1983stochastic} and the Chung-Lu model, with households (CLH)~\cite{ball2009threshold} and without (CL)~\cite{chung2002average}.
The first two models both exhibit clustering commonly found in real-world networks; the CL model was included as a baseline.
For each model, we selected configuration parameters known to produce bounded expansion classes.
In the CL and CLH models, we used a degree distribution with exponential decay and household size four; for SB we used

$$
	\begin{bmatrix}
	.40 & .30 & .20 & .10 \\
	  - & .50 & .13 & .05 \\
	  - &  -  & .35 & .11 \\
	  - &  -  &  -  & .45
	\end{bmatrix}
$$
as the probability matrix.
For each model, three random instances with 1024 vertices and average degree 6 were generated.

\subsection{Hardware}
The experiments were run on identical machines with four-core, 3.0 GHz Intel Xeon E5 v3 processors with a 10 MB cache and 64 GB of memory ($4\times 16$ GB).
Each run of \concuss received dedicated resources to avoid interference from other processes.

    \section{Engineering Evaluation}\label{sec:q12}
We first engineered the implementation of \concuss by testing different configurations against each other before evaluating its practicality via comparison with an existing algorithm.\looseness-1

\subsection{Configuration Testing}\label{sec:config_test}
To measure the run times of various configurations, we created one configuration file for each combination of options described in Sections~\ref{sec:concuss_color} and~\ref{sec:concuss_combine}.
For all possible pairings of random instances and configuration options we ran \concuss three times.
In each run we counted the number of isomorphisms of $P_4$; the particular motif did not vary because the dynamic programming algorithm has theoretically comparable run time for all motifs of the same size.
The metrics of interest in each run were the number of colors used, the total elapsed time, and the time specifically spent in the \textsc{Color} module.
To compare the performance between two implementations of a subroutine, we did a pairwise comparison between configurations in which all other subroutines were held constant.
That is to say, to differentiate between $A_1$ and $A_2$, we would compare $(A_1,B_1,C_1)$ against $(A_2, B_1, C_1)$, $(A_1,B_2, C_1)$ against $(A_2,B_2, C_1)$, etc.

\begin{figure}[!h]
	\centering
	\includegraphics[width=0.5\linewidth]{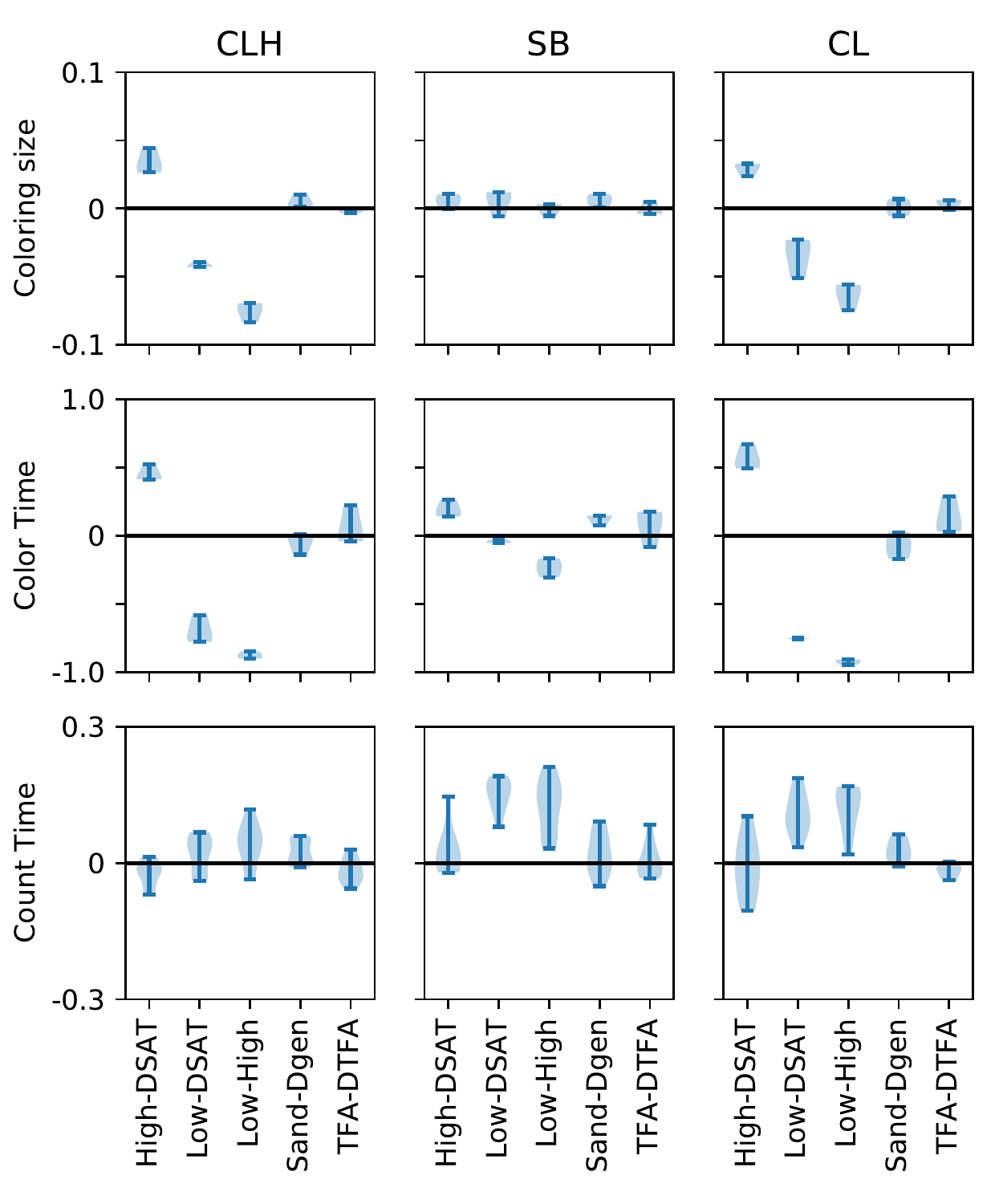}
	\caption{Average difference/sum ratio in coloring size (top), \textsc{Color} time (middle), and \textsc{Decompose}, \textsc{Compute}, and \textsc{Combine} time (bottom) between paired configuration options.  Note the differing $y$-axes.}\label{fig:config_col_diff}
\end{figure}

\begin{figure}[!ht]
	\centering

	\begin{subfigure}[b]{0.5\linewidth}
		\includegraphics[width=0.90\linewidth]{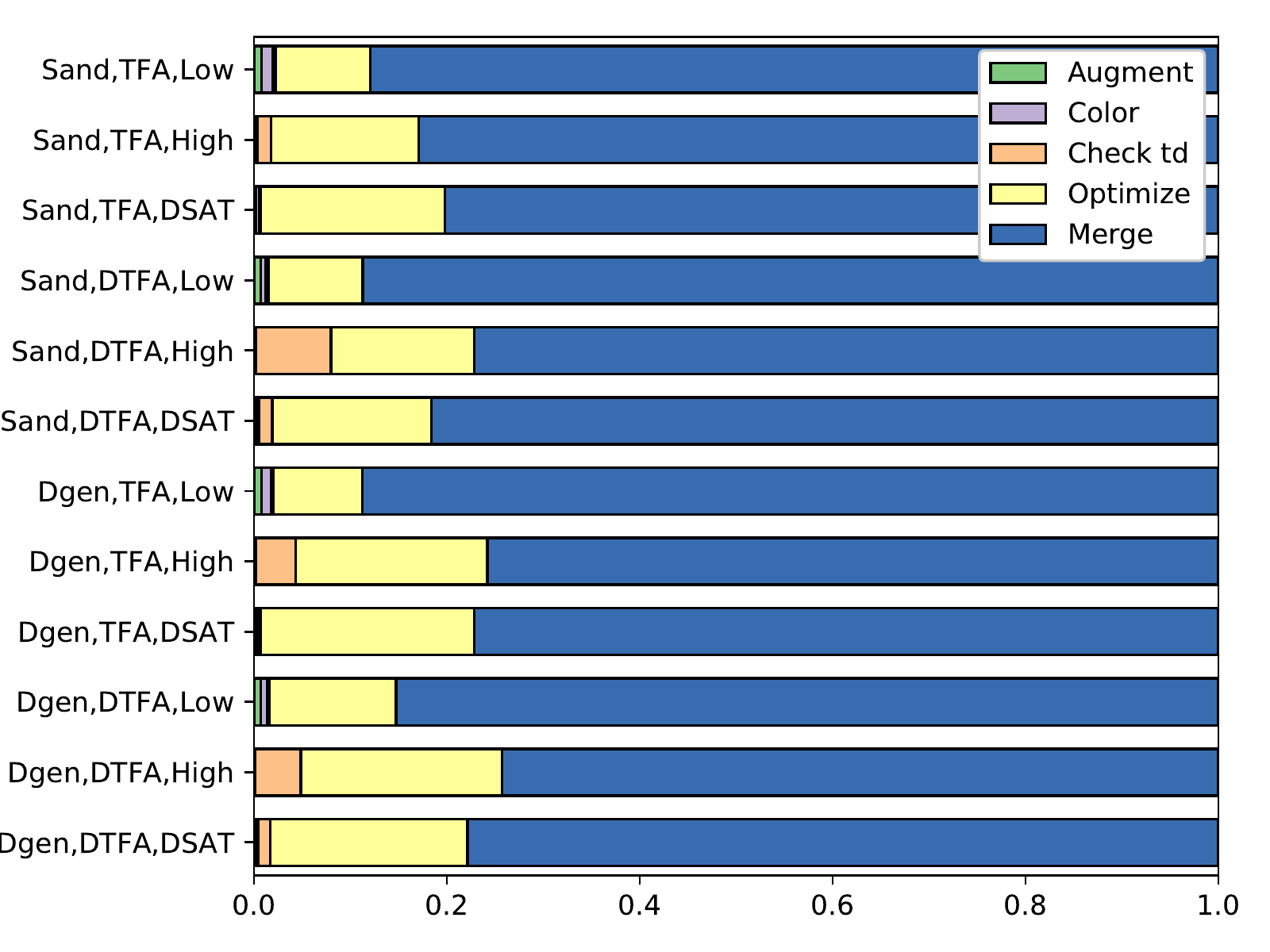}
		\caption{Chung-Lu + households}
	\end{subfigure}%
	\begin{subfigure}[b]{0.5\linewidth}
		\includegraphics[width=0.90\linewidth]{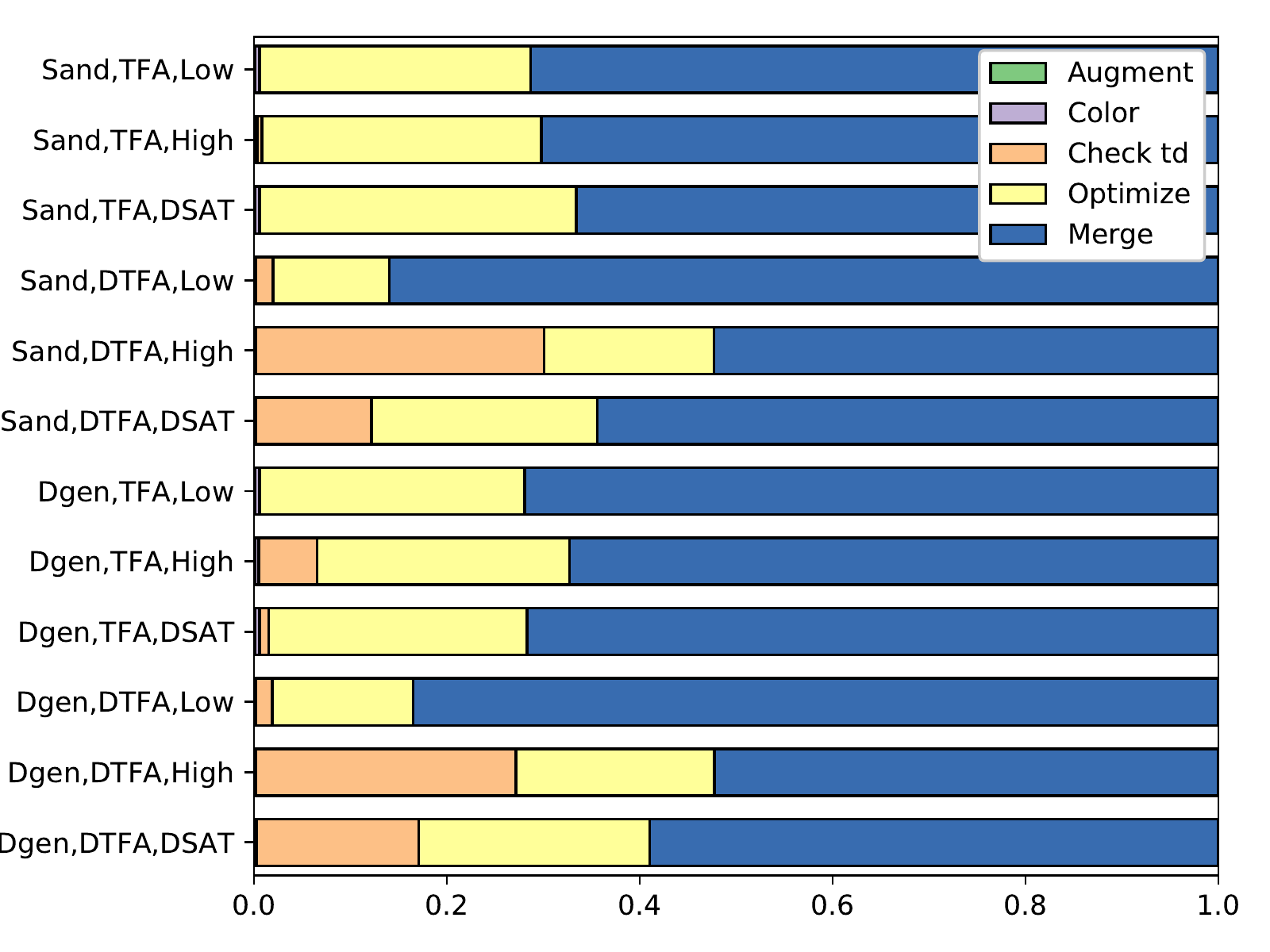}
		\caption{Stochastic block}
	\end{subfigure}
	\caption{Distribution of time spent in the different submodules of the \textsc{Color} module.
	Results for the Chung-Lu model without households were comparable to those for the model with households.}\label{fig:col_dist}
\end{figure}

\begin{figure}[!h]
	\centering
	\includegraphics[width=0.5\linewidth]{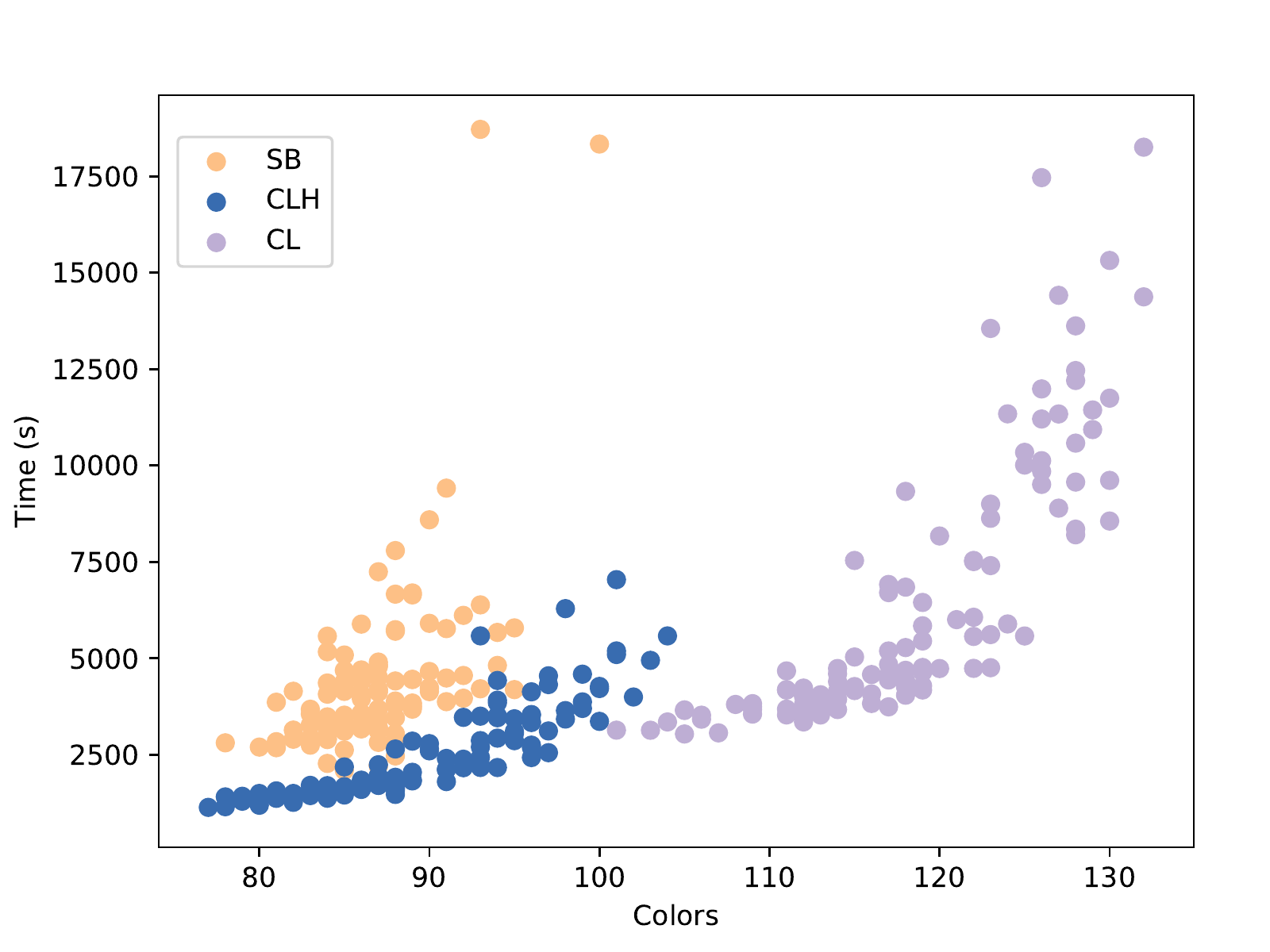}
	\caption{Relationship between number of colors and total execution time using \texttt{Inclusion-exclusion}.}\label{fig:color_reduction}
\end{figure}

Though there were minor variations across graph models, the effects of the different configuration options (Figure~\ref{fig:config_col_diff}) were relatively consistent.
Orienting the edges using \texttt{Sandpiling} and \texttt{Degeneracy} led to comparable time and coloring sizes.
Surprisingly, using \texttt{DTFA}, which is more prudent in augmenting edges, did not outperform \texttt{TFA} in either time or coloring size.
We attribute this result to the fact that the average number of augmentation steps needed to reach a $p$-centered coloring was smaller while using \texttt{TFA} ($2.9$ vs.\ $4.8$), indicating that it may be worth making ``suboptimal'' choices for the augmentations as long as it leads to a $p$-centered coloring sooner.
The \texttt{Low-degree} coloring prioritization generally yielded fewer colors and a shorter run time than \texttt{High-degree}, while the more complicated \texttt{DSATUR} prioritization was somewhere in-between.
Taken as a whole, the best coloring routine used \texttt{Degeneracy} orientation, \texttt{TFA}, and \texttt{Low-degree} prioritization.

The general distribution of time spent within the \textsc{Color} module also did not vary much with the coloring configuration (Figure~\ref{fig:col_dist}).
Computing the augmentations and coloring the vertices took a very small fraction of the total run time;
upwards of 80\% of the time was spent ensuring merging color classes kept the coloring $p$-centered, with the pre-merge optimization consuming most of the rest of the time.
Nearly all of the computation in these two post-processing routines is in deciding whether a smaller coloring is indeed $p$-centered.
As such, small improvements to the efficiency of checking the treedepth of all color sets would pay large dividends in the overall time.\looseness-1

Although the post-processing reduced the coloring size by at least 50\% in 90\% of the colorings and the total execution time correlated positively with coloring size (Figure~\ref{fig:color_reduction}), we cannot assess whether those optimizations were a net benefit without extrapolating beyond our observed data.
Because the time spent merging is dependent on the coloring size, we believe there is a coloring size threshold below which the post-processing is worthwhile; it would be useful to empirically identify this threshold in future work.

The effects of the configurations on the computation downstream from the coloring (Figure~\ref{fig:config_col_diff}) were weaker and more varied.
In particular, we note that \texttt{Low-degree} prioritization often led to slower downstream computation, but this was offset by faster coloring times.\looseness-1

The only configuration option varied outside of the \textsc{Color} module was the method of correcting double counting in the \textsc{Combine} module.
We were once again surprised to see that \texttt{Hybrid}, the ``intelligent'' method of circumventing enumerating many additional color classes, was not successful at improving the run time over the \naive \texttt{Inclusion-exclusion}.
To the contrary, \texttt{Hybrid} increased the total time of the \textsc{Decompose}, \textsc{Compute}, and \textsc{Combine} modules by at least a factor of two, getting up to a factor of ten when the number of colors was very large.
We ultimately concluded the preferred configuration is \texttt{Degeneracy}, \texttt{TFA}, \texttt{Low-degree}, and \texttt{Inclusion-exclusion}.

\subsection{Comparison with \nxvf}
After determining the behavior of various configurations of \concuss, we wanted to assess its performance compared to algorithmic approaches that do not exploit bounded expansion structure.
Recognizing that \concuss is a proof-of-concept of an algorithmic pipeline for classes of bounded expansion, we were primarily interested in establishing it has comparable run times to other algorithms and good scaling behavior;
	in future work, we hope to incorporate additional theoretical advances into \concuss, i.e., those identified in Section~\ref{sec:q34}, to increase its performance.
To make a comparison in the same programming language, we selected the NetworkX~\cite{networkx} implementation of the popular VF2 algorithm~\cite{cordella2004sub}, which we will refer to as \nxvf.
Although there exists an updated version of this algorithm known as VF3~\cite{carletti2017introducing}, the authors' publicly available implementation~\cite{VF3} only accepts graphs with $2^{16}$ or fewer vertices, which is insufficient for the data sets described below.

\subsubsection{Additional Data}
Because of the large $p$-centered coloring sizes of the Chung-Lu and stochastic block model graphs generated for the previous experiment, \concuss requires around 45 minutes to count in graphs where \nxvf terminates in 1 second.
We hypothesized, however, that in large graphs with small coloring sizes and many isomorphisms of $H$ into $G$, \concuss would outperform \nxvf.
We included the restriction on number of isomorphisms because the run time of \concuss does not depend on the count, while \nxvf requires more time for each additional isomorphism.
To test this hypothesis we generated graphs that met the criteria with the following procedure.
First, we generated a complete binary tree of depth $d$.
Then, we selected a tree vertex uniformly at random and attached to it the endpoint of a $P_\ell$.
We continued randomly selecting tree vertices with replacement and adding a $P_\ell$ for a total of $s\cdot 2^{d}$ times.
That is, we made sure that each tree vertex had an average of $s$ $P_\ell$s as neighbors.
We denote this graph $T_{d,s,\ell}$.
In this way, we could effectively vary the size of the graph and the number of isomorphisms of small stars and paths.
\begin{figure}[!h]
	\centering
	\resizebox{0.7\linewidth}{!}{\begin{tikzpicture}[
	vertex/.style={draw, circle, fill=black, minimum size=3mm, inner sep=0pt}
	]
	\node[vertex] (v00) at (3.5,2.0) {};
	\node[vertex] (v10) at (1.5,1.4) {};
	\node[vertex] (v11) at (5.5,1.4) {};
	\node[vertex] (v20) at (0.25,0.8) {};
	\node[vertex] (v21) at (2.75,0.8) {};
	\node[vertex] (v22) at (4.25,0.8) {};
	\node[vertex] (v23) at (6.75,0.8) {};
	\node[vertex] (v30) at (0,0) {};
	\node[vertex] (v31) at (0.5,0) {};
	\node[vertex] (v32) at (2.5,0) {};
	\node[vertex] (v33) at (3,0) {};
	\node[vertex] (v34) at (4,0) {};
	\node[vertex] (v35) at (4.5,0) {};
	\node[vertex] (v36) at (6.5,0) {};
	\node[vertex] (v37) at (7,0) {};

	\draw (v00) edge (v10) edge (v11);
	\draw (v10) edge (v20) edge (v21);
	\draw (v11) edge (v22) edge (v23);
	\draw (v20) edge (v30) edge (v31);
	\draw (v21) edge (v32) edge (v33);
	\draw (v22) edge (v34) edge (v35);
	\draw (v23) edge (v36) edge (v37);

	\node[vertex, fill=black!40] (p00) at (1.2,0.8) {};
	\node[vertex, fill=black!40] (p10) at (1.8,0.8) {};
	\node[vertex, fill=black!40] (p01) at (1.2,0.3) {};
	\node[vertex, fill=black!40] (p11) at (1.8,0.3) {};

	\draw[black!40] (v10) edge (p00) edge (p10);
	\draw[black!40] (p00) edge (p01);
	\draw[black!40] (p10) edge (p11);
	\draw[|-|] (7.5,-0.2) -- node[right] {$d$} (7.5,2.2);
	\draw[|-|] (2.15, 0.1) -- node[right] {$\ell$} (2.15,1.0);
	\draw[|-|] (1,0.0) -- node[below] {$s$} (2, 0.0);

\end{tikzpicture}}
	\caption{$T_{d,s,\ell}$.}\label{fig:xmas_tree}
\end{figure}

\subsubsection{Results}

\begin{figure}[!h]
	\centering
	\resizebox{\linewidth}{!}{\input{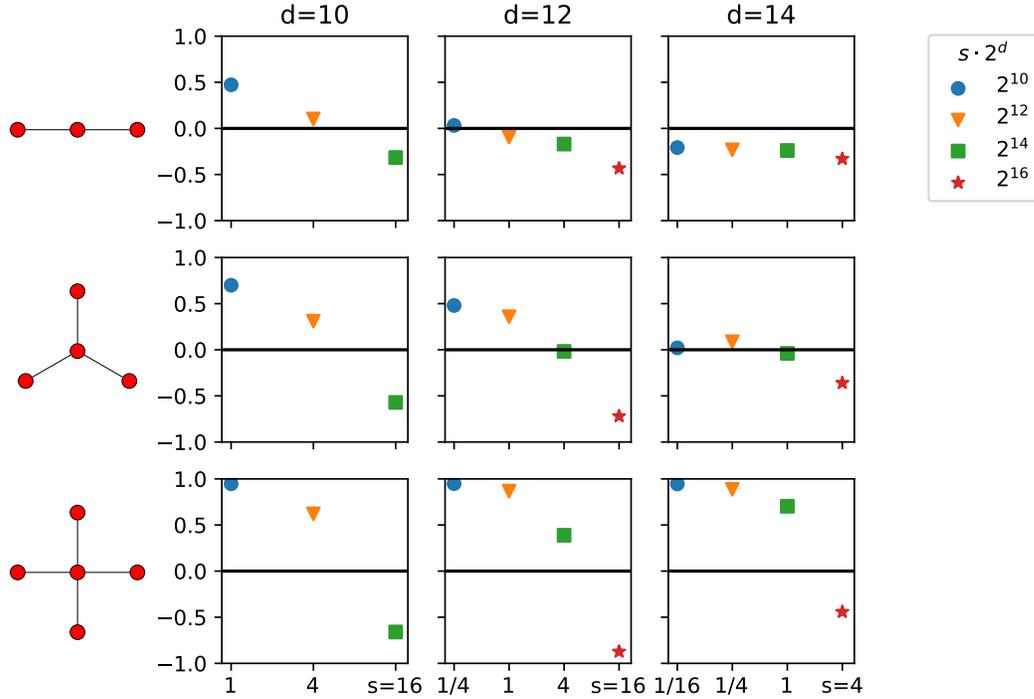}}
	\caption{Average difference/sum ratio between \concuss and \nxvf on $T_{d,s,1}$ as a function of $s$, the average number of $P_1$s per tree vertex.
	Each small plot shows a fixed motif and value of $d$.
	Negative ratios indicate \concuss is faster.}\label{fig:pendant1_times}
\end{figure}

We observed (Figure~\ref{fig:pendant1_times}) that our hypothesis holds when counting stars in $T_{d,s,\ell}$ for sufficiently large $s$.
In particular, \concuss has more than a fourteen-fold speed advantage over \nxvf in counting $S_5$s in $T_{12,16,1}$.
We believe the difference in time would diverge even more for counting $S_5$s in $T_{14,16,1}$, but \nxvf already takes nearly 80 hours in $T_{12,16,1}$.

It is also important to note that when $\ell = 1$, the number of paths does not increase in the same way that the number of stars increases.
This is because every path of length at least four must contain multiple tree vertices, of which each tree vertex is adjacent to at most three.
Moreover, $P_n$ has exactly two automorphisms (``forwards'' and ``backwards'') for every $n$, while $S_n$ has $(n-1)!$ automorphisms (any ordering of the leaves).
We observed that there were insufficiently many $P_4$s and $P_5$s in $T_{d,s,1}$ to see the same performance benefits, and \nxvf was consistently faster than \concuss.

As predicted, \concuss can outperform \nxvf when counting in $T_{d,s,4}$, which has many more $P_4$s and $P_5$s (Figure~\ref{fig:pendant4_times}).
Similar to the results shown in Figure~\ref{fig:pendant1_times} (Appendix~\ref{app:experiment_figures}), \concuss counts $P_3$s faster in all graph sizes, while the advantage in counting $P_5$s is only realized in the largest graphs.

    \section{Bottleneck Identification}\label{sec:q34}
Having identified configurations and graph instances in which \concuss outperforms \nxvf, we proceeded to identify bottlenecks in \concuss that ought to receive further theoretical consideration.
We focused on how changing properties of the output of the \textsc{Color} module would influence performance in the other three modules.

\subsection{Color Class Distribution}
Because a graph may admit many distinct $p$-centered colorings of small size, it is important to know whether two colorings of the same size lead to observable differences in the downstream computation.
Though the effects the coloring has on the rest of the pipeline almost certainly depend on complex and subtle interactions, we hypothesized that the distribution of sizes of color classes is an important factor.
The intuition behind this hypothesis is that different distributions in sizes will lead to different ``shapes'' of treedepth decompositions, i.e., one large, wide decomposition vs.\ many small, thin decompositions.
To test it, we created colorings of equal size that have unequal distributions of color class sizes.\looseness-1

\begin{figure}[!h]
	\centering
	\footnotesize
	\resizebox{0.8\linewidth}{!}{\begin{tikzpicture}[
	every node/.style={draw},
	turnpoint/.style={}
	]

	%
	%

	\coordinate (minsink) at (10.9, 0) {};
	\coordinate (medsink) at (4.5, -2.4) {};
	\coordinate (maxsink) at (10.9, -2.4) {};
	\coordinate (minsource)  at (9.1, -0.0) {};
	\coordinate (medsource)  at (4.5, -1.1) {};
	\coordinate (maxsource)  at (9.1, -1.1) {};
	\draw[->, ultra thick] (minsource) -- node [draw=none, label={\Large \tt min}] {} (minsink);
	\draw[->, ultra thick] (medsource) -- node [draw=none, label=right:{\Large \tt med}] {} (medsink);
	\draw[->, ultra thick] (maxsource) -- node [draw=none, label=below left:{\Large \tt max}] {} (maxsink);

	\node[fill=goodblue,   minimum size=16mm] (1) at (0,0) {};
	\node[fill=goodorange, minimum size=14mm] (2) at (2,0) {};
	\node[fill=goodgreen,  minimum size=12mm] (3) at (4,0) {};
	\node[fill=goodpurple, minimum size=10mm] (4) at (6,0) {};
	\node[fill=goodyellow, minimum size=8mm ] (5) at (8,0) {};
	\draw ($(1.south west)+(-0.2,-0.2)$) rectangle ($(5.north east)+(0.6,0.6)$);

	\begin{scope}[shift={(12,0)}]
		\node[fill=goodblue,   minimum size=16mm] (1) at (0,0) {};
		\node[fill=goodorange, minimum size=14mm] (2) at (2,0) {};
		\node[fill=goodgreen,  minimum size=12mm] (3) at (4,0) {};
		\node[fill=goodpurple, minimum size=10mm] (4) at (6,0) {};
		\node[fill=goodyellow, minimum height=8mm, minimum width=4mm] (5) at (7.7,0) {};
		\node[fill=black!75,   minimum height=8mm, minimum width=4mm] (5') at (8.3,0) {};
		\draw ($(1.south west)+(-0.2,-0.2)$) rectangle ($(5'.north east)+(0.50,0.6)$);
	\end{scope}
	\begin{scope}[shift={(0,-3.5)}]
		\node[fill=goodblue,   minimum size=16mm] (1) at (0,0) {};
		\node[fill=goodorange, minimum size=14mm] (2) at (2,0) {};
		\node[fill=goodgreen,  minimum height=12mm,  minimum width=6mm] (3) at (3.6,0) {};
		\node[fill=black!75,   minimum height=12mm,  minimum width=6mm] (3) at (4.4,0) {};
		\node[fill=goodpurple, minimum size=10mm] (4) at (6,0) {};
		\node[fill=goodyellow, minimum size=8mm ] (5) at (8,0) {};
		\draw ($(1.south west)+(-0.2,-0.2)$) rectangle ($(5.north east)+(0.6,0.6)$);
	\end{scope}
	\begin{scope}[shift={(12,-3.5)}]
		\node[fill=goodblue,   minimum height=16mm, minimum width=8mm] (1) at (-0.5,0) {};
		\node[fill=black!75,   minimum height=16mm, minimum width=8mm] (1') at (0.5,0) {};
		\node[fill=goodorange, minimum size=14mm] (2) at (2,0) {};
		\node[fill=goodgreen,  minimum size=12mm] (3) at (4,0) {};
		\node[fill=goodpurple, minimum size=10mm] (4) at (6,0) {};
		\node[fill=goodyellow, minimum size=8mm ] (5) at (8,0) {};
		\draw ($(1.south west)+(-0.10,-0.2)$) rectangle ($(5.north east)+(0.6,0.6)$);
	\end{scope}

\end{tikzpicture}}
	\caption{Creating a new color class (black) using the three splitting heuristics (\texttt{min}, \texttt{med}, \texttt{max}).}\label{fig:color_class_split}
\end{figure}
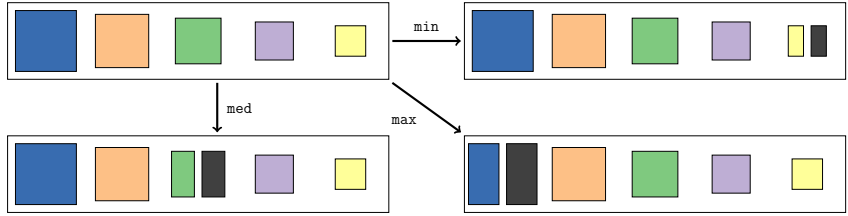

The stochastic block and Chung-Lu with households graphs described previously were used again as inputs for this experiment; we excluded the Chung-Lu graphs without households due to their long run times.
For each graph instance, we found the largest coloring from the configuration experiment, then iteratively split in half the existing color classes (Figure~\ref{fig:color_class_split}) in colorings generated by the other configurations until all colorings had the same size.
To vary the distributions of color class sizes, we used three \emph{splitting heuristics}: \texttt{max}, \texttt{med}, and \texttt{min}, which target the maximum-, median-, and minimum-sized color classes\footnote{Color classes with exactly one vertex were ignored for calculating the minimum and median sizes.}.
When splitting color classes, we chose a random subset of vertices to move into the new color class; to account for this randomness, we repeated each splitting heuristic three times for each coloring.
We ran the \textsc{Decompose}, \textsc{Compute}, and \textsc{Combine} modules of \concuss three times for each graph instance and associated new coloring, using \texttt{Inclusion-Exclusion} in the \textsc{Combine} module.

We observed (Table~\ref{tab:coloring_experiment_result}) that \texttt{max} led to the fastest performance, while \texttt{med} and \texttt{min} had very similar run times.
The dynamic programming algorithm counts isomorphisms from the leaves of the treedepth decomposition upwards using two operations:  \emph{forgetting}, which moves from children to their parent, and \emph{joining} which combines results from ``siblings'' (vertices with the same parent).
Since siblings often belong to the same color class, ensuring that no color class has too many vertices limits the number of children, which in turn limits the number of joins.
In Table~\ref{tab:coloring_experiment_result}, we report that when using \texttt{max} the average number of joins was 33\% lower than when splitting the median or minimum-sized classes.
Thus future coloring algorithms should attempt to make the sizes of the color classes more uniform.
\begin{table}[!h]
	\centering
	\footnotesize
	\begin{tabular}{|c | c | r | r | r |}\hline
	Model & Method & Avg time (s) & Joins & Forgets\\\hline
	SB & \texttt{max} & 856 & \num{5.12e05} & \num{2.00e08} \\\hline
	SB & \texttt{med} & +19\% & +48\% & +13\% \\\hline
	SB & \texttt{min} & +20\% & +49\% & +13\% \\\hline
	CLH & \texttt{max} & 1327 & \num{7.63e05} & \num{3.01e08} \\\hline
	CLH & \texttt{med} & +21\% & +48\% & +4\% \\\hline
	CLH & \texttt{min} & +21\% & +50\% & +5\% \\\hline
\end{tabular}

	\caption{Average run time and dynamic programming operations used for each color splitting heuristic.
	 The statistics for \texttt{med} and \texttt{min} are reported as percent increases over \texttt{max}.}\label{tab:coloring_experiment_result}
\end{table}

\subsection{Color Set Treedepth}
The dynamic programming (DP) in the \textsc{Compute} module counts isomorphisms of a subgraph of size $h$ in an arbitrary treedepth decomposition of depth $t$.
While this algorithm runs in linear time with respect to the size of the graph, it takes $O(t^h)$ time with respect to $t$.
One potential direction for further theoretical improvements to the bounded expansion pipeline is to find alternatives to $p$-centered colorings in which subgraphs with at most $i<p$ colors have treedepth that is small but greater than $i$.
Ideally, relaxing treedepth requirement would result in a significant reduction in the size of the coloring without dramatically increasing the time spent in DP in the \textsc{Compute} module.
Though we cannot measure the size of a hypothetical coloring, we can evaluate how the time per DP operation increases as we count the same subgraph in larger treedepth decompositions.
To do this, we computed a $6$-centered coloring, enumerated sets of $3$, $4$, and $5$ colors, and counted the number of isomorphisms of $P_3$ in each set.\looseness-1

The run time of the DP algorithm is only dependent on the treedepth of the decompositions insofar as it increases the number of labelings\footnote{These are the $k$-patterns mentioned in Appendix~\ref{sec:concuss_counting}; see~\cite{demaine2014structural}.} of the ancestors of each vertex.
There are $t^3+3t^2+t$ such labelings for a $P_3$, each of which requires a constant number of DP operations, so the amount of time spent counting in one color set should grow proportionally to this function.

\begin{figure}[!h]
	\centering
	{\includegraphics[width=0.4\linewidth]{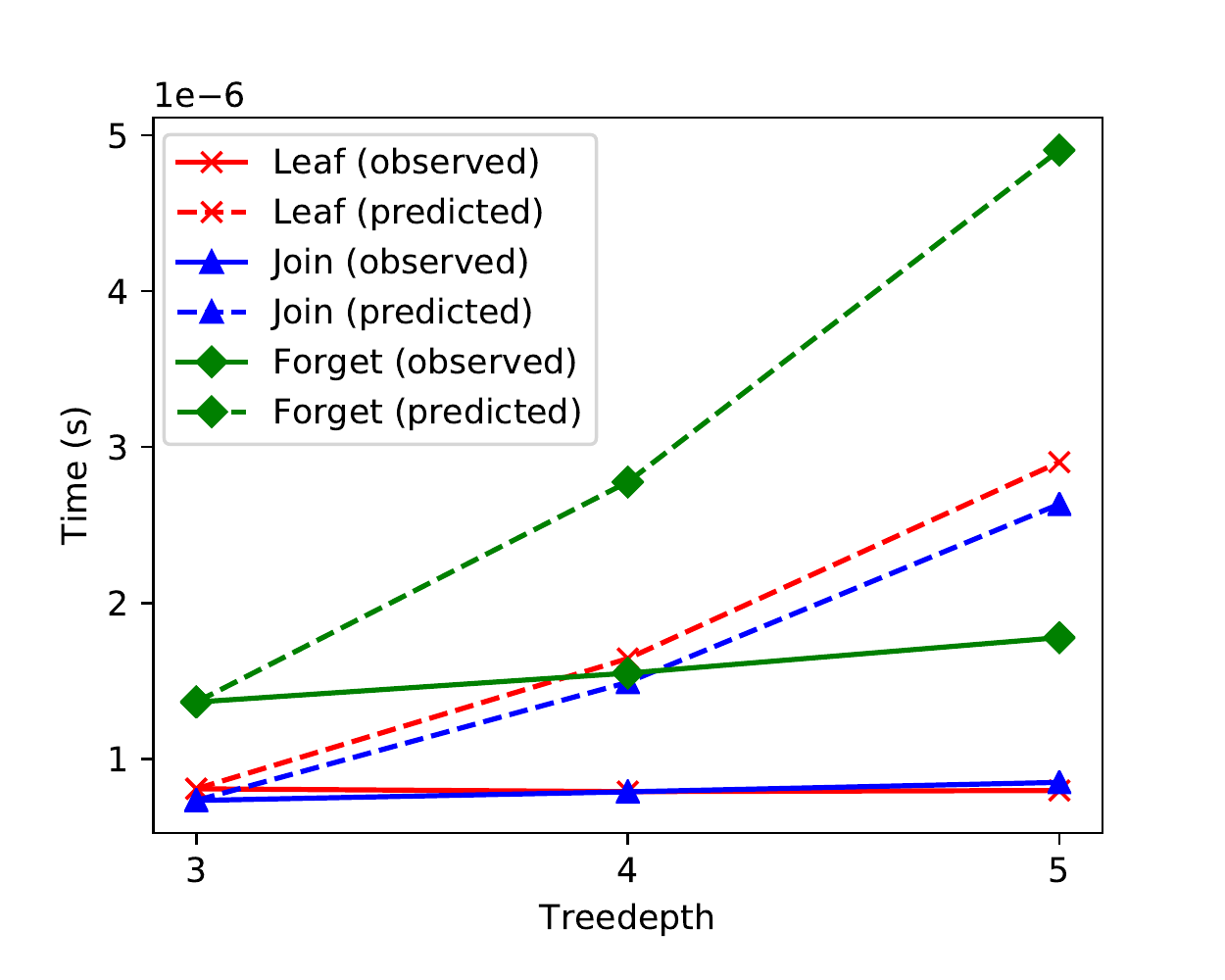}}
	\caption{Observed vs.\ predicted average execution time per operation in the stochastic block graphs.
		Results for the graphs from the CL models are similar and are omitted for space.
		We only include those operations incurred in counting in color sets of exactly $t$ colors.}\label{fig:td_theory_time}
\end{figure}

\begin{table}[!h]
	\centering
	\footnotesize
	\begin{tabular}{|c|c|c|c|c|c|}\hline
Model & $t$ & Time per & Time per & Time per & Avg.\\
 & & leaf & join & forget & depth\\\hline
	CL & 3 & \num{7.5e-07} & \num{7.3e-07} & \num{1.4e-06} & 2.0\\\hline
	CL & 4 & \num{7.7e-07} & \num{8.1e-07} & \num{1.7e-06} & 2.4\\\hline
	CL & 5 & \num{7.9e-07} & \num{9.0e-07} & \num{2.0e-06} & 2.8\\\hline
	CLH & 3 & \num{8.0e-07} & \num{7.4e-07} & \num{1.4e-06} & 2.0\\\hline
	CLH & 4 & \num{8.0e-07} & \num{8.1e-07} & \num{1.7e-06} & 2.4\\\hline
	CLH & 5 & \num{8.1e-07} & \num{9.3e-07} & \num{2.1e-06} & 2.8\\\hline
	SB & 3 & \num{8.1e-07} & \num{7.3e-07} & \num{1.4e-06} & 1.9\\\hline
	SB & 4 & \num{7.9e-07} & \num{7.9e-07} & \num{1.5e-06} & 2.3\\\hline
	SB & 5 & \num{8.0e-07} & \num{8.5e-07} & \num{1.8e-06} & 2.7\\\hline
\end{tabular}

	\caption{Average execution time per operation and average depth of vertices in the treedepth decomposition.
		Rows for treedepth $t$ only include those operations incurred in counting in color sets of exactly $t$ colors.}\label{tab:td_jf_stats}
\end{table}

To compare our observations to the predicted growth of $t^3+3t^2+t$, we assumed that the average time per operation for $t=3$ aligned correctly with this theoretical time and extrapolated outwards for $t\in \{4,5\}$.
As reported in Figure~\ref{fig:td_theory_time}, the observed times grow significantly slower than predicted for all operations.
The time spent on each operation is dependent on the depth of the vertex in the treedepth decomposition, but the average depth of vertices does not grow at a one-to-one ratio with the treedepth (Table~\ref{tab:td_jf_stats}).
This means that the additional color classes add vertices at many depths of the decomposition, rather that simply increasing the depth of every branch uniformly, and thus we seldom ``pay'' for the cost of the larger depths.
Consequently, we conclude that colorings producing deeper decompositions with fewer colors ought to be a viable strategy for improving the overall speed of \concuss.\looseness-1

    \section{Conclusion}
We demonstrated that exploiting bounded expansion structure is a promising methodology for scaling subgraph isomorphism counting to larger classes of sparse graphs.
We identified through testing various configurations of \concuss that orienting with a degeneracy ordering, augmenting with TFA, coloring low-degree vertices first, and combining counts using the inclusion-exclusion principle led to fast run times and small colorings.
On sufficiently sparse graphs with high isomorphism counts, \concuss was able to significantly outperform \nxvf.
In order to get the bounded expansion pipeline to outperform other methods in a broader set of graphs, it is important to understand in future work whether the large coloring sizes of the random graph models were an artifact of the heuristics used in generating the $p$-centered colorings, or whether their asymptotically bounded values are actually large.
Moreover, reimplementing \concuss in a lower-level language like C++ would allow us to make meaningful comparisons to other C++ implementations, e.g., the RI algorithm in~\cite{leskovec2016snap}.

Our experiments also identified three areas for future theoretical research.
The first is to reduce the time needed to verify a coloring is $p$-centered, which would yield large dividends in the \textsc{Color} module.
The second is to design $p$-centered coloring algorithms that balance the sizes of the color classes, consequently reducing the number of join operations.
Finally, our results suggest using colorings that trade fewer colors for larger treedepth per color set could be faster than using $p$-centered colorings.\looseness-1

\section*{Acknowledgments}
{\small This work was supported in part by the DARPA GRAPHS Program and the
Gordon \& Betty Moore Foundation's Data-Driven Discovery Initiative through
Grants SPAWAR-N66001-14-1-4063 and GBMF4560 to Blair D.~Sullivan.}

    \bibliographystyle{plain}
    \bibliography{be,subgraphIsomorphism}
    \pagebreak
    \appendix
\section{Computing Treedepth Decompositions}\label{app:tdd}

\begin{algorithm}[!h]
	\caption{Treedepth decomposition implied by centered coloring}
	\label{alg:cc_to_tdd}
	\begin{algorithmic}
		\STATE $\phi \leftarrow$ centered coloring of $G$
		\STATE $T\leftarrow $ treedepth decomposition of $G$
		\IF{$|G|=0$}
			\RETURN
		\ENDIF
		\STATE $v \leftarrow \operatorname{center}(\phi|_{G})$
		\STATE $\operatorname{root}(T) \leftarrow v$
		\FORALL{components $C\in G\backslash \{v\}$}
			\STATE $T'\leftarrow $ treedepth decomposition of $C$ implied by $\phi|_{C}$
			\STATE $\operatorname{parent}(\operatorname{root}(T')) \leftarrow v$
		\ENDFOR
	\end{algorithmic}
\end{algorithm}

\section{Algorithmic Description of \concuss}\label{app:concuss}

\subsection{Color}\label{sec:concuss_color}
\begin{algorithm}[!h]
	\caption{Sketch of $p$-centered coloring}
	\label{alg:p-centered}
	\begin{algorithmic}
		\STATE $G'\leftarrow G$
		\REPEAT
			\STATE \emph{Orient} the edges of $G'$ to make it a directed acyclic graph of low indegree
			\STATE \emph{Augment} $G'$ with additional edges based on their orientation
			\STATE \emph{Color} the vertices greedily
			\STATE \emph{Check} whether each set of vertices with $p-1$ colors has low treedepth in $G$
		\UNTIL{the coloring is $p$-centered}
	\end{algorithmic}
\end{algorithm}
The workflow in the \textsc{Color} module is outlined in Algorithm~\ref{alg:p-centered}.
The objective is to produce a $p$-centered coloring by greedily coloring the vertices of the graph.
Because a simple greedy coloring may not immediately be $p$-centered, additional edges are added to the graph to impose further constraints.
After a bounded number of iterations of the loop in Algorithm~\ref{alg:p-centered}, a greedy coloring is guaranteed to be $p$-centered~\cite{boundedexpansion1,boundedexpansion3,FelixThesis}.
Since each substep runs in linear time with respect to $|G|$, the entire \textsc{Color} module also runs in linear time.

\textit{Orient}:
The later stages of the \textsc{Color} module operate on directed acyclic graphs.
In essence, directed edges capture the ancestor relationships in treedepth decompositions of small sets of colors by having edges point from an ancestor to its descendants.
In this representation, vertices with large indegree have many ancestors, indicating subgraphs with large treedepth.
A low-indegree orientation can be found either with a \texttt{Degeneracy} ordering~\cite{matula1983smallest} or by \texttt{Sandpiling}~\cite{bak1987self}.

\textit{Augment}:
After the edges are oriented, additional directed edges are augmented to the graph to impose more constraints.
The placement of the augmenting edges is dependent on the direction of edges in the orientation.
Two vertices $a,b$ are said to be \emph{transitive} if there is some vertex $c$ such that $ac$ and $cb$ are both directed edges in the graph, while $a,b$ are said to be \emph{fraternal} if there is a vertex $c$ such that $ac$ and $bc$ are both directed edges.
The \emph{transitive-fraternal augmentation} (\texttt{TFA}) algorithm~\cite{boundedexpansion2} adds the edge $ab$ for every pair of transitive vertices $a,b$ and either the edge $xy$ or $yx$ for every pair of fraternal vertices $x,y$.
The orientation of edges between fraternal vertices is chosen to preserve acyclicity and minimize indegree.
\emph{Distance-truncated transitive fraternal augmentation} (\texttt{DTFA})~\cite{FelixThesis} works similarly, but only requires that some transitive and fraternal edges are added.
If $ac$ was added in the $i$th iteration\footnote{Edges present originally in $G$ are considered to have been added in the 1st iteration.} of the loop in Algorithm~\ref{alg:p-centered} and $cb$ was added in the $j$th, the edge $ab$ need not be added until the $(i+j)$th iteration.

\textit{Color}:
The acyclic orientation of the edges defines an ordering over which to greedily color the vertices.
To minimize the number of colors, \concuss uses of one of three different heuristics:  prioritizing high-degree vertices (\texttt{High-degree}), prioritizing low-degree vertices (\texttt{Low-degree}), or prioritizing vertices with many colors already represented among their neighbors (\texttt{DSATUR})~\cite{brelaz1979new}.

\textit{Check}:
Finally, we check whether the coloring is $p$-centered by testing whether each connected subgraph that uses $p-1$ or fewer colors has a unique color.
This can be done by keeping track of components and their respective color multiplicities using union-find structures; when two components merge, we ensure that that there is still a unique color.
We perform this check at every iteration in order to prevent the algorithm from adding any more colors than necessary.

\textit{Other Optimizations}:
We include additional pre- and post-processing steps to obtain smaller colorings.
Vertices of high degree can cause many additional edges to be augmented at each iteration of the loop, which in turn causes more colors to be used.
For this reason, removing all high-degree vertices, e.g., degree at least $\sqrt[4]{n}$, finding a coloring of the smaller graph, and then giving each removed vertex a new unique color can reduce the total number of colors.
Likewise, in any component of size larger than two, vertices of degree one can be removed and all given the same color in post-processing.
After obtaining a $p$-centered coloring, we can also potentially reduce the number of colors by randomly adding transitive and fraternal edges, computing a new greedy coloring, and seeing if this greedy coloring is $p$-centered.
The final post-processing step is to check whether pairs of color classes can be merged without violating the $p$-centered property.

\subsection{Decompose}
After the coloring is computed, \concuss finds the components induced by each set of $|H|$ colors.
To save computation, \concuss processes overlapping color sets sequentially in a depth-first search manner.
For example, if $|H| = 4$ and the coloring has 9 colors, the sets would be processed as:  \dots, \{1, 2, 3\}, \{1, 2, 3, 4\}, \{1, 2, 3, 5\}, \dots, \{1, 2, 3, 9\}, \{1, 2, 4\}, \{1, 2, 4, 5\}, etc.
When processing color set \{1, 2, 3, 4\}, the components from \{1, 2, 3\} are saved to be reused for \{1, 2, 3, 5\}.

Given a component induced by a particular color set, a treedepth decomposition is created using Algorithm~\ref{alg:cc_to_tdd}.
In order to save memory, these treedepth decompositions are fed into the subsequent \textsc{Count} and \textsc{Compute} stages before a new color set is processed.

\subsection{Compute}\label{sec:concuss_counting}
The \textsc{Compute} module uses the dynamic programming algorithm of Demaine et al.~\cite{demaine2014structural} to count the isomorphisms of $H$ in a treedepth decomposition.
It has exponential time dependence on the small treedepth, but runs in linear time with respect to $|G|$.
Briefly, this algorithm exploits the fact that no edges in the treedepth decomposition join vertices in different branches.
Therefore, partial pieces of $H$ in one subtree have limited ways of interacting with those in other subtrees.
By tracking which pieces occur in each subtree, these counts of partial pieces can be combined to count all isomorphisms in the decomposition.

Integral to the dynamic programming are \emph{$k$-patterns}, also called boundaried graphs.
These (sub)graphs have a subset of vertices labeled uniquely with integers on the interval $[1,k]$, which in turn determine how partial pieces of $H$ can ``glue'' together.
We store the $k$-pattern labelings as bitvectors to enable fast checking of whether two patterns are compatible to be ``glued'', i.e., their labelings are consistent with one another.

\subsection{Combine}\label{sec:concuss_combine}
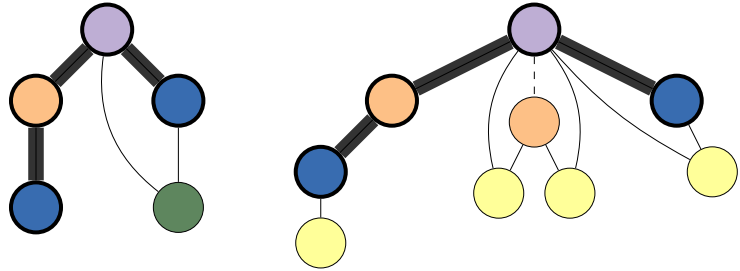
\begin{figure}[!h]
	\centering
	\resizebox{0.7\linewidth}{!}{\begin{tikzpicture}
	\node[vertex,fill=goodblue,ultra thick] (A) at (-3,0.5) {};
	\node[vertex,fill=goodorange,ultra thick] (B1) at (-3,2) {};
	\node[vertex,fill=goodpurple,ultra thick] (C) at (-2,3) {};
	\node[vertex,fill=goodblue,ultra thick] (B2) at (-1,2) {};
	\node[vertex,fill=goodgreen!60!black!90] (D) at (-1,0.5) {};

	\foreach \from/\to in {A/B1,B1/C,C/B2}
	\draw[double distance=3pt, double=black!80,black!80,ultra thick](\from) -- (\to);

	\foreach \from/\to in {A/B1,B1/C,C/B2,B2/D}
	\draw(\from) -- (\to);

	\foreach \from/\to in {C/D}
	\draw[](\from) to [bend right] (\to);

	\node[vertex,fill=goodpurple,ultra thick] (a1) at (4,3) {};
	\node[vertex,fill=goodorange,ultra thick] (b1) at (2,2) {};
	\node[vertex,fill=goodorange] (b2) at (4,1.7) {};
	\node[vertex,fill=goodblue,ultra thick] (b3) at (6,2) {};
	\node[vertex,fill=goodblue,ultra thick] (c1) at (1,1) {};
	\node[vertex,fill=goodyellow] (c2) at (3.5,0.7) {};
	\node[vertex,fill=goodyellow] (c3) at (4.5,0.7) {};
	\node[vertex,fill=goodyellow] (c4) at (6.5,1) {};
	\node[vertex,fill=goodyellow] (d1) at (1,0) {};

	\foreach \from/\to in {a1/b1,b1/c1,a1/b3}
	\draw[double distance=3pt, double=black!80,black!80,ultra thick](\from) -- (\to);

	\draw[] (a1) to (b1);
	\draw[] (a1) to (b3);
	\draw[] (a1) to [bend right=25] (c2);
	\draw[] (a1) to [bend left=25] (c3);
	\draw[] (a1) to [bend right=10] (c4);
	\draw[] (b1) to (c1);
	\draw[] (b2) to (c2);
	\draw[] (b2) to (c3);
	\draw[] (b3) to (c4);
	\draw[] (c1) to (d1);
	\draw[dashed] (a1) to (b2);

\end{tikzpicture}}
	\caption{The highlighted $P_4$ only uses three colors and thus will be counted in multiple color sets of size four.}\label{fig:double_count}
\end{figure}

The goal in the \textsc{Combine} module is to sum the counts from each treedepth decomposition to get the count for the whole graph.
However, because the color sets contain overlapping subgraphs, a simple summation of counts may count some isomorphisms more than once.
For example, if $H = P_4$, the subgraph in Figure~\ref{fig:double_count} will be counted in the color sets \{blue, orange, purple, yellow\}, \{blue, orange, purple, green\}, etc.
There are multiple ways to rectify this issue.
\texttt{Inclusion-exclusion} simply enumerates all color sets with fewer than $|H|$ colors, counts the isomorphisms in those subgraphs, and then applies the inclusion-exclusion principle to correct the counts appropriately.
We can also incorporate the combinations into the dynamic programming itself by ignoring counts of color sets that have already been seen (\texttt{Hybrid}).
\vfill
\pagebreak
\section{Additional Figures}\label{app:experiment_figures}

\begin{figure}[!h]
	\centering
	\resizebox{\linewidth}{!}{\input{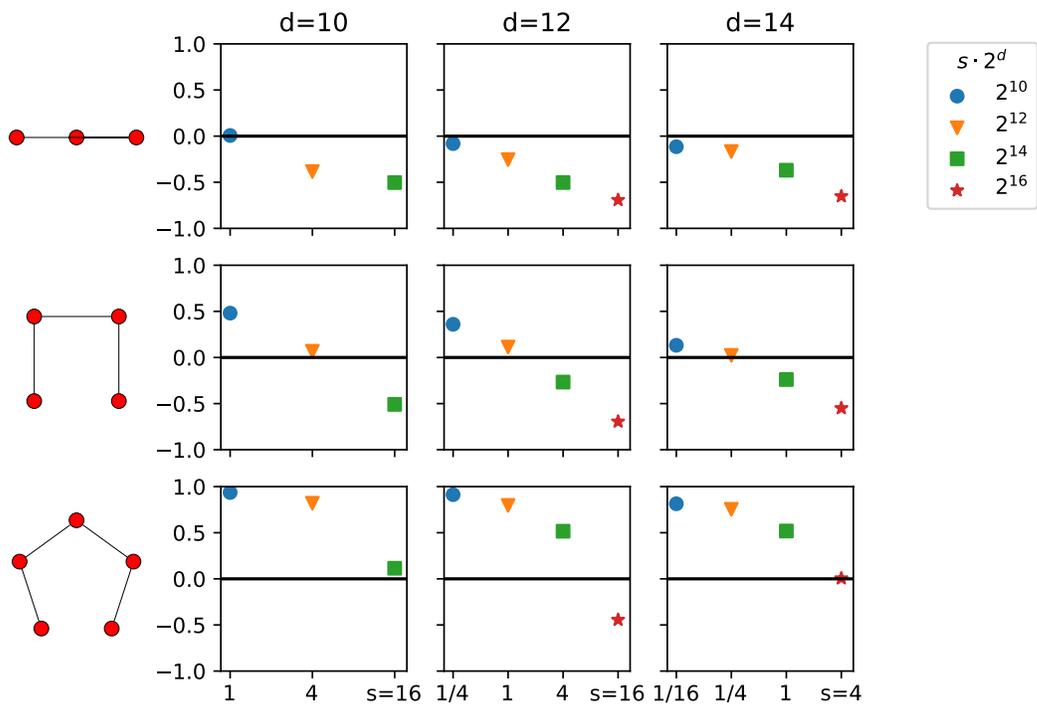}}
	\caption{Average difference/sum ratio between \concuss and \nxvf on $T_{d,s,4}$ as a function of $s$, the average number of $P_4$s per tree vertex.
	Each small plot shows a fixed motif and value of $d$.
	Negative ratios indicate \concuss is faster.}\label{fig:pendant4_times}
\end{figure}

\end{document}